\documentclass[twocolumn,aps,floatfix,showpacs]{revtex4}
\usepackage[dvips]{graphicx}
\begin{document}
\title{Polaron in t-J model}
\author{
A.S.~Mishchenko$^{1,2}$ and N.~Nagaosa$^{1,3}$}
\affiliation{$^1$CREST, Japan Science and Technology Agency (JST), 
AIST, 1-1-1, Higashi, Tsukuba 305-8562, Japan 
\\ 
$^2$RRC ``Kurchatov Institute", 123182, Moscow, Russia 
\\
$^3$CREST, Department of Applied Physics, The University of Tokyo, 7-3-1
Hongo, Bunkyo-ku, Tokyo 113, Japan }

\begin{abstract}
We present numeric results for ground state and angle resolved 
photoemission spectra (ARPES) for single hole in t-J model coupled 
to optical phonons. 
The systematic-error free diagrammatic Monte Carlo is employed where the 
Feynman graphs for the Matsubara Green function in imaginary time are 
summed up completely with respect to phonons variables, while magnetic 
variables are subjected to non-crossing approximation.  
We obtain that at electron-phonon coupling constants relevant for high $T_c$   
cuprates the polaron undergoes self-trapping crossover to strong 
coupling limit and theoretical ARPES demonstrate features observed 
in experiment: a broad peak in the bottom of the spectra has momentum 
dependence which coincides with that of hole in pure t-J model. 
\end{abstract}


\maketitle

In the context of broad interest to the challenging properties of the 
high-$T_c$ superconductors, a single hole in the 
Mott insulator has been studied extensively in terms of the t-J model
\begin{equation}
\hat{H}_{\mbox{\scriptsize t-J}} = - t \sum_{\langle ij \rangle s} 
                  c_{is}^{\dagger} c_{js}
                + J \sum_{\langle ij \rangle} 
\left(
{\bf S}_i {\bf S}_j - n_i n_j / 4
\right) ,
\label{tJ}
\end{equation}
where $c_{j\sigma}$ is projected (to avoid double occupancy) fermion 
annihilation operator, $n_i < 2$ is the occupation number, ${\bf S}_i$ is
spin-1/2 operator, and $\langle ij \rangle$ denotes nearest-neighbour 
sites in two-dimensional lattice.
Different theoretical approaches (for review and recent studies see 
Refs.~\onlinecite{kane,ManoRev94,Assad00,Mis01}) give
consistent results for the Lehman Spectral Function (LSF)
$L_{\bf k}(\omega) = - \pi^{-1} \Im G_{\bf k}(\omega)$ of the Green 
function $G_{\bf k}(\omega)$. 
Namely, the LSFs at all momenta have a quasiparticle (QP) peak in the 
low energy part together with a broad incoherent continuum extending 
up to the energy scale of the order of $t$. 
Sharp QP peak in the LSF of the ground state at momentum 
${\bf k}=(\pi/2,\pi/2)$ has the weight $Z \sim J/t$. 
This QP peak is sharp at all momenta and it's energy 
dispersion has the bandwidth $W_{J/t} \sim J$. 
More advanced t-$t'$-$t''$-J model takes into account long range hopping
amplitudes $t'$ and $t''$ which alter the bandwidth and dispersion of the 
QP resonance 
\cite{Zaanen_95,Xiang_96,Kyu_Fer,tklee1,tklee2,Tohyama_00,Shen_03}.
It was shown \cite{Zaanen_95} that, depending on the parameters of the 
t-$t'$-$t''$-J model, the QP peak can be either enhanced or 
completely suppressed at a large part of the Brillouin zone. 
However, at parameters which are needed for description of ARPES measurements 
on carefully studied \cite{ZX95} Sr$_2$CuO$_2$Cl$_2$ 
($J/t \approx 0.4$, $t'/t \approx -0.3$, see \cite{Xiang_96,Kyu_Fer,tklee1}), 
the QP peak remains well defined for all momenta \cite{Zaanen_95,Yunoki,MNSUP}.
Possible damping of the peak \cite{Zaanen_95}, in contrast to experiment,
is much less than the QP bandwidth. 
Therefore, the properties of QP peak in t-$t'$-$t''$-J model
at realistic parameters are the same as in generic t-J model.     
 
Experimentally, ARPES in the undoped cuprates revealed the LSF of a 
single hole \cite{Tohyama_00,Shen_03}
and observed dispersion of the lowest peak in the LSF is in good 
agreement with the theoretical predictions of the t-$t'$-$t''$-J model
\cite{Xiang_96,Kyu_Fer,tklee1,tklee2,Tohyama_00,Shen_03}.  
The puzzling point is that, in contrast to the theory, experiments never 
show  sharp QP resonance and a broad peak with the width of 
the order of 0.1-0.5eV ($\approx t$) is observed instead. 
Note, that broadening is seen just in undoped systems where the ground 
state of the single hole is the lowest energy state in the Hilbert space
of the (N-1)-electron problem. One can rule out a possibility of an 
extrinsic origin of this width since the doping introduces further 
disorder, while a sharper peak is observed in overdoped region. 

The role of electron-phonon (e-ph) coupling has gained the recent renewed 
interest. 
One reason is that the ARPES data in doped metallic cuprates revealed the 
energy dispersion strongly renormalized by e-ph interaction \cite{Shen_03}.
Besides, the strong e-ph interaction is crucial for explanation
of renormalization and lineshapes of phonons observed in neutron 
scattering experiments \cite{Khal0107,Khal0108}. 
In addition, the large isotope effect on $T_c$ for underdoped cuprates and 
on the superfluid density at the optimal doping suggests the 
vital role of e-ph coupling \cite{FromPNN1}.

Strong and intermediate coupling regimes of e-ph interaction 
in t-J model can not be studied neither by exact diagonalization 
\cite{Fehske_98} on small clusters nor by self consistent Born 
approximation (SCBA) where both magnon and phonon propagators are 
subjected to non crossing approximation (NCA)
which neglects the vertex corrections 
\cite{Ramsak_92,Kyung_96,Yunoki}.  
Small system size implies a discrete spectrum and, hence, neither 
crossover from large to small size polaron nor the problem of 
linewidth in ARPES can be addressed in the former approach.   
The latter method omits the Feynman diagrams with mutual crossing of 
phonon propagators and, hence, neglects the vortex corrections which are 
crucial for treating the e-ph interaction in the strong coupling regime (SCR). 
One can use the NCA for the interaction of the hole with magnetic system 
because  spin S=1/2 can not flip more than one time and magnon cloud around 
the hole saturates. 
To the contrary, phonon-phonon non-crossing approximation (PPNCA) fails for 
SCR since number of phonon quanta around the hole is not limited.  

The key role of the vortex corrections for e-ph interaction in SCR can 
be demonstrated by numerically exact Diagrammatic Monte Carlo (DMC) method 
\cite{PS,MPSS,MN01,Aus02} where Feynman graphs for the Matsubara Green 
function of a hole in phonon bath are generated by Monte Carlo and 
summed up without systematic errors. 
In the framework of this technique one can compare results of
complete summation of Feynman expansion with those of restricted summation 
where diagrams with inter-crossing of phonon propagators are neglected. 
In order to extract the effect of vortex corrections on results 
we considered conventional two dimensional Holstein model, where
the complete and PPNCA summation of Feynman expansion by DMC can be 
performed exactly.
In this model hole freely hops with the amplitude $t$ ($t$ is set to 
unity below) and interacts with dispersionless (frequency 
$\Omega=\mbox{const}$) optical phonons  by short range coupling $\gamma$   
\begin{equation}
\hat{H}^{\mbox{\scriptsize e-ph}} = 
\Omega \sum_{\bf k} b_{\bf k}^{\dagger} b_{\bf k}
+
N^{-1} \gamma \sum_{\bf k , q}  
\left[ h_{\bf k}^{\dagger} h_{\bf k-q} b_{\bf k} + h.c.
\right] \;.
\label{e-ph}
\end{equation}  
Exact and PPNCA results for energy and $Z$-factor of polaron are in good 
agreement at small values $g \le 0.2$ of dimensionless interaction constant 
$g=\gamma^2/(8t\Omega)$ while are crucially different in the intermediate and 
strong coupling regimes. 
E.g., for $\Omega/t=0.1$ exact result shows crossover to SCR at 
$g > g_{H}^{c} \approx 0.6$ while in PPNCA the polaron 
is in weak coupling regime even for $g=60$. 
We conclude that DMC is the only method which can treat intermediate 
and strong coupling regime of the e-ph interaction for the 
problem of one hole in the macroscopic system.

In this Letter we present a study of a single hole in t-J model  
interacting with dispersionless optical phonons (\ref{e-ph}) by DMC 
\cite{PS,MPSS,MN01,Aus02} method.
We found  that due to slowing down of the hole by spin flip cloud around it,
the hole in t-J model is subject to the stronger 
e-ph coupling than the freely propagating hole and hence undergoes 
the crossover to SCR at smaller coupling. This is in contrast to 
the naive expectation that the small $Z$ factor reduces the e-ph coupling
in t-J model.
Besides, we found that SCR occurs at e-ph couplings which are typical for 
high $T_c$ materials. 
Finally, our results for SCR qualitatively reproduce data of 
ARPES experiments: a {\it broad} peak, whose 
{\it energy dispersion is similar to that of pure} t-J {\it model}, 
dominates in low energy part of LSF. 

In the standard spin-wave approximation in momentum representation 
\cite{Liu_92} the dispersionless hole $\varepsilon_0 = const$ 
(annihilation operator is $h_{\bf k}$) propagates
in the magnon (annihilation operator is $\alpha_{\bf k}$) bath
\begin{equation}
\hat{H}_{\mbox{\scriptsize t-J}}^{0} =
\sum_{\bf k} \varepsilon_0 h_{\bf k}^{\dagger} h_{\bf k} 
+
\sum_{\bf k} \omega_{\bf k} \alpha_{\bf k}^{\dagger} \alpha_{\bf k}
\label{h0}
\end{equation}
with magnon dispersion $\omega_{\bf k}=2J\sqrt{1-\gamma_{\bf k}^2}$, where
$\gamma_{\bf k}=(\cos k_x + \cos k_y) / 2$. The hole is scattered by magnons 
\begin{equation}
\hat{H}_{\mbox{\scriptsize t-J}}^{\mbox{\scriptsize h-m}} =
N^{-1} \sum_{\bf k , q} M_{\bf k , q} 
\left[ h_{\bf k}^{\dagger} h_{\bf k-q} \alpha_{\bf k} + h.c.
\right] 
\label{h-m}
\end{equation}  
with the scattering vortex 
$M_{\bf k , q} = 4t(u_{\bf q}\gamma_{\bf k-q} + v_{\bf q} \gamma_{\bf k})$,
where $u_{\bf k} = \sqrt{(1+\nu_{\bf k})/(2\nu_{\bf k})}$, 
$v_{\bf k} = - sign (\gamma_{\bf k})\sqrt{(1-\nu_{\bf k})/(2\nu_{\bf k})}$,
and $\nu_{\bf k}=\sqrt{1-\gamma_{\bf k}^2}$. 
We chose the value $J/t=0.3$ at which NCA for magnons is shown to be 
sufficiently good approximation \cite{Liu_92} and set the phonon 
frequency $\Omega=0.1$ to make it similar to experimental value.  

We generate Feynman expansion of Matsubara Green function of a hole in 
momentum representation for the  infinite system at zero temperature 
by DMC \cite{MPSS,MN01,Aus02}.
Then we obtain LSF by stochastic optimization method \cite{MPSS,RP} which  
resolves equally well both sharp and broad features of the spectra. 
Diagrams with crossing of a magnon propagator by both magnon and phonon 
lines are neglected \cite{sign} though the diagrams with 
inter-crossing of phonon propagators are taken into account. 
The DMC algorithm for pure t-J model eqs.\ (\ref{h0}-\ref{h-m})
is equivalent to the macroscopic limit of SCBA approach on finite 
lattices \cite{Liu_92}. 
When PPNCA is introduced to DMC at finite couplings to phonons, the algorithm
is nothing but the thermodynamic limit of SCBA approach on finite lattices
to the complete model (\ref{e-ph}-\ref{h-m}) when both magnons and phonons are
taken into account in NCA \cite{Kyung_96}.
Comparison of our data with results of Refs.~\cite{Liu_92,Kyung_96} shows that
$Z$-factor and energy of the lowest peak are weakly influenced by finite 
size corrections since the relative discrepancy is less than $10^{-2}$.
The shapes of LSF in SCBA approach are very similar to those obtained by 
DMC. We observed only slight discrepancy in widths and energies of high energy 
peaks.
Therefore, the main advantage of DMC over other existing methods is 
the possibility to take into account phonon-phonon vertex corrections in
macroscopic system.  

%%%%%%%%%%%%%%%%%%%%%%%%%%%%%%%%%%%%%%%%%
\begin{figure}[bht]
\hspace{-0.45 cm}  \vspace {-0.5 cm}
\includegraphics{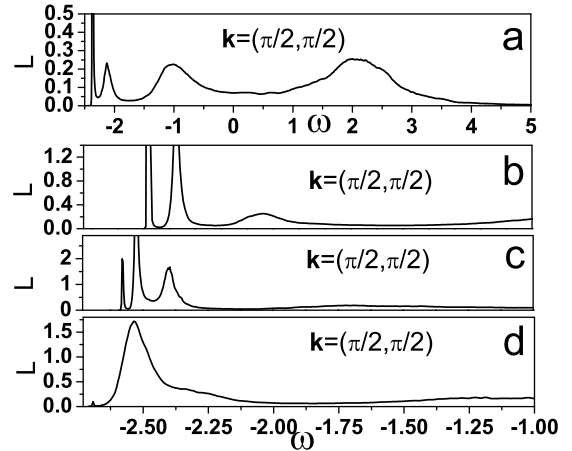}
\caption{\label{fig:fig1} Hole LSF in ground state at $J/t=0.3$:
(a) for $g=0$; (b-d) low energy part for g=0.1445 [$\gamma=0.34$] (b),
g=0.2 [$\gamma=0.2$] (c), and g=0.231125 [$\gamma=0.43$] (d). } 
\end{figure}
%%%%%%%%%%%%%%%%%%%%%%%%%%%%%%%%%%%%%%%%%

%%%%%%%%%%%%%%%%%%%%%%%%%%%%%%%%%%%%%%%%%
\begin{figure}[htb]
\hspace{-0.3 cm}  \vspace {-0.5 cm}
\includegraphics{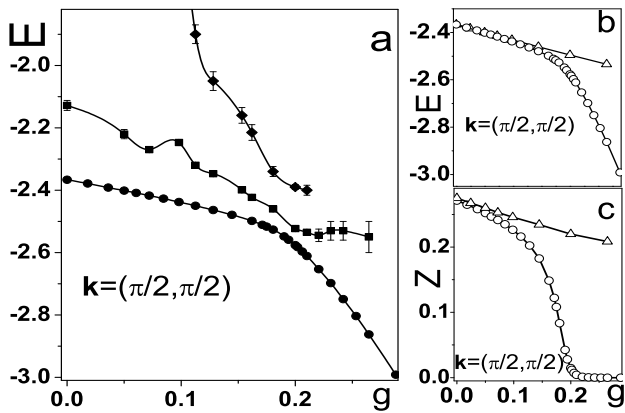}
\caption{\label{fig:fig2} Dependence on coupling strength at 
$J/t=0.3$:
(a) energies of lowest LSF resonances;
(b [c]) energy [$Z$-factor] of lowest peak in 
DMC (circles) and PPNCA (triangles). 
Lines are to guide an eye. The errorbar, if not shown, is 
less than the point size.}
\end{figure}
%%%%%%%%%%%%%%%%%%%%%%%%%%%%%%%%%%%%%%%%%

%%%%%%%%%%%%%%%%%%%%%%%%%%%%%%%%%%%%%%%%%
\begin{figure}[bht]
\hspace{-0.3 cm}  \vspace {-0.5 cm}
\includegraphics{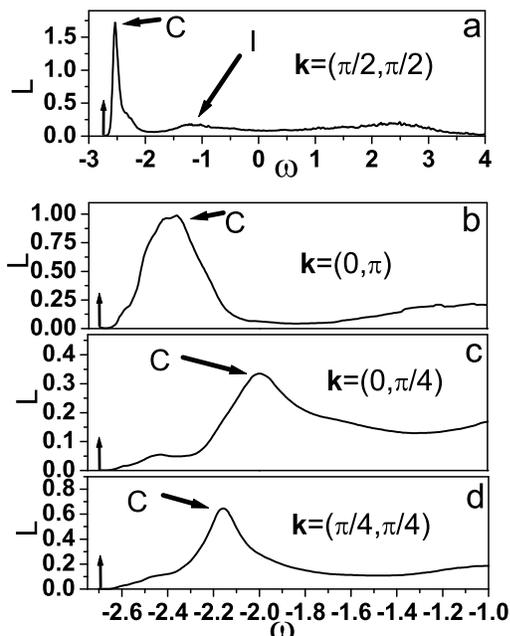}
\caption{\label{fig:fig3} LSF of the hole at 
$J/t=0.3$ and $g=0.231125$:
(a) full energy range for ${\bf k}=(\pi / 2, \pi / 2)$;
(b-d) low energy part for different momenta.
Slanted arrows show broad peaks which can be interpreted in ARPES
spectra  as coherent (C) and incoherent (I) part.
Vertical arrows indicate position of ground state resonance which
is not seen in the vertical scale of the figure.}
\end{figure}
%%%%%%%%%%%%%%%%%%%%%%%%%%%%%%%%%%%%%%%%%

%%%%%%%%%%%%%%%%%%%%%%%%%%%%%%%%%%%%%%%%%
\begin{figure}[tbh]
\hspace{-0.3 cm}  \vspace {-0.5 cm}
\includegraphics{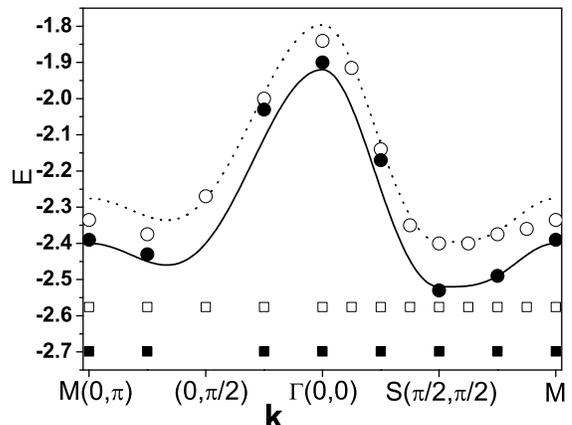}
\caption{\label{fig:fig4} Dispersion of resonances energies
at $J/t=0.3$: 
broad resonance (filled circles) and lowest polaron resonance 
(filled squares) at $g=0.231125$; third broad resonance 
(open circles) and lowest polaron resonance (open squares)
at $g=0.2$.
The solid curves are dispersions (\ref{Marsig}) of a hole in pure 
t-J model at $J/t=0.3$ ($W_{J/t=0.3}=0.6$):
$\varepsilon_{min}=-2.396$ ($\varepsilon_{min}=-2.52$) for solid
(dotted) line.}
\end{figure}
%%%%%%%%%%%%%%%%%%%%%%%%%%%%%%%%%%%%%%%%%

Figure \ref{fig:fig1} presents our results for LSF in the ground state. 
Three low energy peaks for $g=0$ (Fig.~\ref{fig:fig1}a) are string resonances 
\cite{Shrain88} since their energies, as we found in calculations of LSF for
different exchange constants $0.1 \le J \le 0.4$, obey the scaling low 
$a+b_i(J/t)^{2/3}$ ($a=-3.13$, $b_{1,2,3}=4.89/2.09/1.63$). 
With increasing of $g$ these peaks are observed up to $g=0.21$ 
(Figs.~\ref{fig:fig1}b and \ref{fig:fig1}c) while the peak with highest 
energy broadens and disappears at higher couplings (Fig.~\ref{fig:fig1}d).

Dependence of the peak energies (Fig.~\ref{fig:fig2}a) and ground state
$Z$-factor (Fig.~\ref{fig:fig2}c) on $g$ resembles picture inherent in 
self-trapping phenomenon \cite{Rashba82,RP}: the states cross and hybridize 
at critical coupling  $g_{\mbox{\scriptsize t-J}}^{c} \approx 0.19$ 
and $Z$-factor of ground state resonance rapidly decreases.
We note, that PPNCA result does not show transition into SCR even at 
considerably larger $g$ (see Fig.~\ref{fig:fig2}b and Fig.~\ref{fig:fig2}c).  
According to general understanding of self-trapping crossover,
at small $g$ the ground state is weakly coupled to phonons while excited 
resonances have strong lattice deformation.
At critical coupling the crossing and hybridization of these states occurs and 
for higher $g$ the roles of these states exchange: the lowest 
state is strongly trapped by lattice deformation while the upper ones are 
nearly free. 
Therefore, above the crossover point $g > g_{\mbox{\scriptsize t-J}}^{c}$ 
one expects that the lowest state should be dispersionless while the upper 
resonances have to show considerable momentum dependence.
This assumption is supported by momentum dependence of LSF well above the
crossover point (Fig.~\ref{fig:fig3}).  
The energy of the lowest peak is momentum independent while the bandwidth
of the upper broad resonance (Fig.~\ref{fig:fig4}) is the same 
($W_{J/t=0.3}=0.6$) as it is in the pure t-J model. 
The most surprising peculiarity of the momentum dependence of broad 
resonance is that it is exactly the same as expected for t-J model 
with no coupling to phonons: it obeys the scaling relation 
\begin{eqnarray}
\varepsilon_{\bf k} & = & \varepsilon_{min} +
W_{J/t}
\{
[\cos k_x + \cos k_y]^2 / 5 +
\nonumber 
\\ 
& &
[\cos (k_x+k_y) + \cos (k_x-k_y)]^2/4 
\},
\label{Marsig}
\end{eqnarray}     
which describes dispersion for pure t-J model in the broad range of 
exchange constants $0.1 \le J/t \ge 0.9$ \cite{Marsiglio91}. 
We emphasize, that unrenormalized dispersion of the upper resonances 
is the general property of SCR. E.g., for the coupling $g=0.2$, which is
slightly higher than the critical coupling $g_{\mbox{\scriptsize t-J}}^{c}$, 
(see Fig.~\ref{fig:fig1}c and \ref{fig:fig2}), dispersion of the upper 
resonance also obeys eqs. (\ref{Marsig}) while the lowest peak is momentum 
independent (see Fig.~\ref{fig:fig4}). 

The behavior of LSF in SCR is exactly the same as it is observed in ARPES
experiments. The LSF consists of {\it broad QP} peak and 
high-energy incoherent continuum (see Fig.~\ref{fig:fig3}). Besides, 
momentum dependence of the broad QP peak is similar to dispersion 
of sharp resonance in the pure t-J model 
(see Fig.~\ref{fig:fig4}). 
The lowest lying dispersionless peak in SCR has small $Z$-factor and can 
not be discerned in ARPES experiments: it`s spectral weight is small
($Z \sim 0.015$) at $g=0.2$ (Fig.~\ref{fig:fig1}c) and completely suppressed 
($Z<10^{-3}$) at $g=0.231125$ (Fig.~\ref{fig:fig1}d).
On the other hand, the momentum dependence of spectral weight $Z'$ of broad 
dispersive resonance in SCR is akin to that in pure t-J model. For 
$g=0.231125$ the weights of broad peak at ${\bf k}=(\pi/2,\pi/2)$
($\sim 0.27$) and at ${\bf k}=(0,0)$ 
($\sim 0.05$) coincide with $Z$-factors of sharp resonances at
corresponding momenta in pure t-J model.  
Our calculation show that this similarity is the robust feature of SCR: 
$Z'_{(\pi/2,\pi/2)} \sim 0.27 \pm 0.01$ and $Z'_{(0,0)} \sim 0.05 \pm 0.005$
in the broad range of coupling constants $0.21 < g < 0.27$.

Comparing the critical interaction 
$g_{\mbox{\scriptsize t-J}}^{c} \approx 0.19$ for the hole 
in t-J model and critical coupling $g_{H}^{c} \approx 0.6$ for Holstein
model with the same value of hopping $t$, we conclude that interaction with
spins enhances e-ph coupling and accelerates transition into SCR. 
Emission of magnons on each hopping shrinks
the bandwidth of lowest lying resonance by the factor of $J/t$ while the 
reduction of e-ph vertex by the factor $Z$ seems to be absent. This makes the  
influence of e-ph interaction on low energy part of LSF more effective. 
We emphasize that the critical interaction for transition to SCR
is low enough for t-J model to bring the realistic system into SCR. 
E.g., theoretical estimate of the interaction strength from the fitting 
of phonon energies and lineshapes to neutron scattering experiment  gives 
rather large magnitude for e-ph coupling in La(Sr)CuO$_4$ 
\cite{Khal0107,Khal0108}. 
Since the interaction vertex is momentum dependent in the model 
\cite{Khal0107,Khal0108}, we can establish only lower and upper bound 
for effective coupling $g$ in our model with short-range interaction.  
The averaging over the Brillouin zone gives the lower bound
$g_{\mbox{\scriptsize exp}}^{\mbox{\scriptsize min}} \ge 0.15$  
while the maximal value at the Brillouin zone boundary determines 
the upper bound 
$g_{\mbox{\scriptsize exp}}^{\mbox{\scriptsize max}} \le 0.5$. 
However, since self-trapping is governed mainly by short-range coupling 
\cite{Rashba82}, effective constant in our model is more close to the 
upper bound 
$g_{\mbox{\scriptsize exp}}^{\mbox{\scriptsize max}}$. 
Thus, realistic high $T_c$ cuprates are expected to be in SCR.

Finally, we conclude that puzzling behavior of ARPES spectra in 
undoped high $T_c$ materials, which manifests oneself in unexpectedly 
{\it broad} quasiparticle peak with dispersion corresponding to pure 
Mott insulator model, is driven by strong electron-phonon interaction.

Fruitful discussions with N.\ V.\ Prokof'ev, B.\ V.\ Svistunov, 
G.\ Khaliullin and G.\ A.\ Sawatzky are acknowledged. 
This work was supported by RFBR 01-02-16508.


\begin{thebibliography}{99}
\bibitem{kane} C.\ L. Kane, P.\ A.\ Lee, and N.\ Read, 
         Phys.\ Rev.\ B {\bf 39}, 6880 (1989).  
%
\bibitem{ManoRev94} E.\ Manousakis, Rev.\ Mod.\ Phys.\ {\bf 63}, 1 (1991);
         E.\ Dagotto, {\it ibid}.\ {\bf 66}, 763 (1994). 
% 
\bibitem{Assad00} M.\ Brunner, F.\ F.\ Assaad, and A.\ Muramatsu, 
         Phys.\ Rev.\ B {\bf 62}, 15480 (2000).
%
\bibitem{Mis01} A.\ S.\ Mishchenko, N.\ V.\ Prokof\'ev, and B.\ V.\ 
         Svistunov, Phys.\ Rev.\ B {\bf 64}, 033101 (2001).
%
\bibitem{Zaanen_95} J.\ Ba\l a, A.\ M.\ Ole\'{s}, and J.\ Zaanen, 
         Phys.\ Rev.\ B {\bf 52}, 4597 (1995). 
%
\bibitem{Xiang_96} T.\ Xiang and M.\ Wheatley, Phys.\ Rev.\ B {\bf 54},
         R12653 (1996).
%
\bibitem{Kyu_Fer} B.\ Kyung and R.\ A.\ Ferrell, Phys.\ Rev.\ B {\bf 54},
         10125 (1996).
%
\bibitem{tklee1} T.\ K.\ Lee and C.\ T.\ Shih, 
         Phys.\ Rev.\ B {\bf 55}, 5983 (1997).
% 
\bibitem{tklee2} T.\ K.\ Lee, C-M.\ Ho, and N.\ Nagaosa,
         Phys.\ Rev.\ Lett.{\bf 90}, 067001 (2003). 
%
\bibitem{Tohyama_00} T.\ Tohyama and S.\ Maekawa, Superconductors 
         Science and Technology {\bf 13}, R17 (2000).
%
\bibitem{Shen_03} A.\ Danmascelli, Z.-X.\ Shen, and Z.\ Hussain,
         Rev.\ Mod.\ Phys.\ {\bf 75}, 473 (2003).
%
\bibitem{ZX95} B.\ O.\ Wells, Z.-X.\ Shen, A.\ Matsuura, D.\ M.\ King,
         M.\ A.\ Kastner, M.\ Greven, and R.\ G.\ Birgenau,
         Phys.\ Rev.\ Lett.{\bf 74}, 964 (1995).  
%
\bibitem{Yunoki} S.\ Yunoki, A.\ Macridin, and G.\ A.\ Sawatzky, private 
         communication, unpublished
%
\bibitem{MNSUP} A.\ S.\ Mishchenko and N.\ Nagaosa, unpublished.
%
\bibitem{Khal0107} P.\ Horsch and G.\ Khaliullin, in {\it Open Problems in
         Strongly Correlated Electron Systems}, J.\ Bonca et.\ al.\ (eds.),
         Kluwer-Academic, Boston, p.\ 81 (2001). 
%
\bibitem{Khal0108} P.\ Horsch, G.\ Khaliullin, and V.\ Oudovenko, 
         Physica C {\bf 341-348}, 117 (2000).
%
\bibitem{FromPNN1} J.\ Hofer, K. Conder, T.\ Sasagawa, Guo-meng Zhao,
         M.\ Willemin, H. Keller, and K. Kishio, Phys. Rev. Lett.\ {\bf 84},
         4192 (2000).
%
\bibitem{Fehske_98} B.\ Bauml, G.\ Wellein, and H.\ Fehske, Phys.\ Rev.\ 
         B {\bf 58}, 3663 (1998).
%
\bibitem{Ramsak_92} A.\ Ramsak, P.\ Horsch, and P.\ Fulde, Phys.\ Rev.\ 
         B {\bf 46}, 14305 (1992).
%
\bibitem{Kyung_96} B.\ Kyung, S.\ I.\ Mukhin, V.\ N.\ Kostur, and 
         R.\ A.\ Ferrel, Phys.\ Rev.\ B {\bf 54}, 13167 (1996).
%
\bibitem{PS} N.\ V.\ Prokof'ev and B.\ V.\ Svistunov, Phys.\ Rev.\ Lett.\
         {\bf 81}, 2514 (1998). 
%
\bibitem{MPSS} A.S.\ Mishchenko, N.V.\ Prokof'ev, A.\ Sakamoto,
             and B.V.\ Svistunov, Phys. Rev. B {\bf 62}, 6317 (2000).
%
\bibitem{MN01} A.\ S.\ Mishchenko and N.\ Nagaosa, Phys.\ Rev.\ Lett.\ 
         {\bf 86}, 4624 (2001).
%
\bibitem{Aus02} A.\ S.\ Mishchenko, N.\ Nagaosa, N.\ V.\ Prokof'ev, 
         B.\ V.\ Svistunov, and E.\ A.\ Burovskii,
         J.\ Nonlinear Opt.\ {\bf 29}, 257 (2002).
%
\bibitem{Liu_92} Z.\ Liu and E.\ Manousakis, Phys.\ Rev.\ 
         B {\bf 45}, 2425 (1992).
%
\bibitem{RP} A.S.\ Mishchenko, N.\ Nagaosa, N.V.\ Prokof'ev, A.\ Sakamoto,
             and B.V.\ Svistunov, Phys. Rev. B {\bf 66}, 020301 (2002).
%
\bibitem{sign} To avoid sign problem arising from the momentum dependence of 
               magnon scattering vortex $M_{\bf k , q}$.
%
\bibitem{Shrain88} B.\ I.\ Shraiman and E.\ Sigga, Phys.\ Rev.\ Lett.\
             {\bf 60}, 740 (1988).
%
\bibitem{Rashba82} E.\ I.\ Rashba, {\it Self-Trapping of Excitons}, in
         {\it Modern Problems in Condensed Matter Sciences}, Edited by
         V.\ M.\ Agranovich and A.\ A.\ Maradudin (North-Holland, 
         Amsterdam, 1982), Vol.\ 2, p.\ 543. 
%
\bibitem{Marsiglio91} F.\ Marsiglio, A.\ E.\ Ruckenstein, 
         S.\ Schmitt-Rink and C.\ M.\ Varma, Phys.\ Rev.\ B {\bf 43},
         10882 (1991).
\end{thebibliography}
\end{document}